\documentclass[12pt]{iopart}

\usepackage{iopams}  
\usepackage{graphicx}

\begin{document}

\title[Simplification of Biochemical Completion Times]{Simplicity of Completion Time Distributions for Common Complex Biochemical Processes}

\author{Golan Bel$^{1,2,\dagger}$, 
Brian Munsky$^{1,2,\ast}$, 
Ilya Nemenman$^{1,\ddagger}$}

\address{$^1$ Center for Nonlinear Studies and the Computer, Computational, and Statistical Sciences Division,
Los Alamos National Laboratory,
 Los Alamos, NM 87545. USA}
\address{$^2$ Contributed Equally}
\address{$\dagger$ E-mail: golanbel@gmail.com}
\address{$\ast$ E-mail: brian.munsky@gmail.com}
\address{$\ddagger$ E-mail: ilya@menem.com}

\begin{abstract}
Biochemical processes typically involve huge numbers of individual
reversible steps, each with its own dynamical rate constants. For
example, kinetic proofreading processes rely upon numerous sequential
reactions in order to guarantee the precise construction of specific
macromolecules. In this work, we study the transient properties of
such systems and fully characterize their first passage (completion)
time distributions. In particular, we provide explicit expressions for
the mean and the variance of the completion time for a kinetic
proofreading process and computational analyses for more complicated
biochemical systems. We find that, for a wide range of parameters, as
the system size grows, the completion time behavior simplifies: it
becomes either deterministic or exponentially distributed, with a very
narrow transition between the two regimes. In both regimes, the
dynamical complexity of the full system is trivial compared to its
apparent structural complexity. Similar simplicity is likely to arise
in the dynamics of many complex multistep biochemical processes. In
particular, these findings suggest not only that one may not be able
to understand individual elementary reactions from macroscopic
observations, but also that such understanding may be unnecessary.

\end{abstract}

\maketitle

\section{Introduction}
Considering the ever increasing quantity of known
biochemical reactions, one cannot help but be amazed and daunted by
the incredible complexity of the implied cellular networks.  For
example, just a handful of different proteins can form a
combinatorially large number of interacting molecular species, such as
in the case of immune signaling \cite{faeder-03}, where multiple
receptor modification sites result in a model with 354 distinct
chemical species. One must then ask: When do all details of this
seemingly incomprehensible complexity actually matter, and when is
there a smaller set of aggregate, coarse-grained dynamical variables,
parameters, and reactions that approximate the salient features of the
system's dynamics? What determines which features are relevant and
which are not? And if the networks have a simple equivalent dynamics,
did nature choose to make them so complex in order to fulfill a
specific biological function? Or is the unnecessary complexity a
``fossil record'' of the evolutionary heritage?

\begin{figure}[!ht]
\begin{center}
\includegraphics[width=8.7cm]{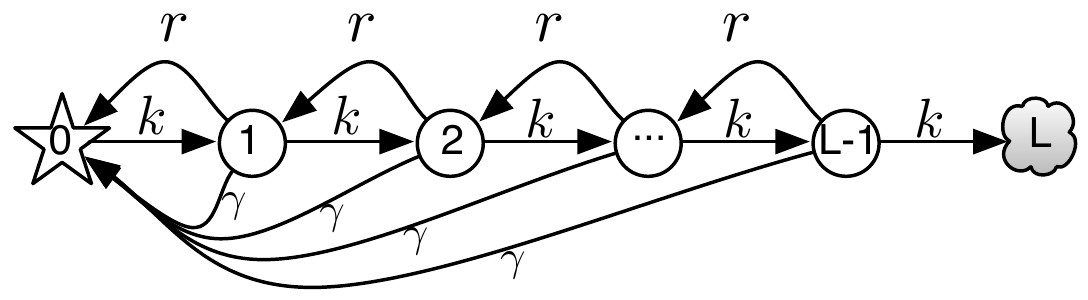}
\end{center}
\caption{{\bf Schematic description of the model.} The process begins
 at the site $i=0$, represented with a star.  At each site, the process
 may transition one step to the right with the forward rate $k$, one step
 to the left with the backward rate $r$, or all the way back to the origin
 with the return rate $\gamma$. The right-most site, $i=L$ is an absorbing site
 (cloud) at which the process is completed.}\label{fig:Model}
\end{figure}

In this article, we begin investigation of these questions in the
context of certain biochemical kinetics networks, namely a reversible
linear pathway, a {kinetic proofreading} (KPR) scheme
\cite{Hopfield:1974}, their combination, and an extension to a much
more arbitrary multistep completion process. These motifs are common
in a variety of cellular processes--including DNA synthesis and repair
\cite{Yan:1999,Sancar:2004}, protein translation
\cite{Hopfield:1974,Blanchard:2004}, molecular transport \cite{NPC},
receptor-initiated signaling
\cite{McKeithan:1995,Rabinowitz:1996,Rosette:2001,Liu:2001,Goldstein:2004,Hlavacek:2002},
and other processes--where assembly of large biochemical structures
requires multiple reversible steps. However, in this article, we leave
aside the functional behavior of these networks and focus instead on a
different question: do these complex kinetic schemes have a
simplified, yet accurate description? Since multistep structural
complexity (see Fig.~\ref{fig:Model}) is crucial for kinetic
proofreading, the KPR process is an ideally suited example for this
analysis, but our conclusions will extend to numerous other complex
biochemical processes.

We show analytically and numerically that, over broad ranges of
parameters, different kinetic schemes exhibit the behavior of either a
deterministic process, or a single-step exponential-waiting-time
process. We also propose intuitive arguments for the result, which
leads us to believe that similar simplifications of complex behavior
may be wide-spread, and even universal. We support this conjecture by
numerically studying a few more complex systems, but leave a general
mathematical proof of this conjecture to future work.

\subsection{The Model}
For this study we begin with a general KPR (gKPR) model
\cite{Hopfield:1974}, for which many properties can be computed
analytically. The model is represented by the Markov chain in Fig.\
\ref{fig:Model}.  At time $t=0$, the dynamics begins at the point
represented by the star ($i=0$).  The process can leave this state at
some exponentially distributed waiting time, defined by a {\em forward
  rate} $k$, and the process can continue in the forward direction
with rate $k$ until it reaches the final absorbing point (cloud) at
$i=L$.  At each interior point, $i\in\{1,2,\ldots,L-1\},$ the process
can also move one step to the left with a {\em backward rate} $r$ or
all the way back to the origin with a {\em return} or {\em
  proofreading rate} $\gamma$. The forward and the backward rates
emphasize the reversibility of all reactions, and the return rate
corresponds to a catastrophic failure, after which the whole process
must start anew. For example, in immune signaling, $\gamma$ would
represent the rate of receptor-ligand dissociation, which destroys
receptor cross-linking and prevents future forward events for a
relatively long period of time \cite{faeder-03}.

This model is substantially simplified compared to detailed models of
real biological processes \cite{faeder-03} in that, in nature, all
three rates may depend on $i$, and the nodes may not form a single
linear chain.  Even so, the detailed understanding of this simplified
model provides an excellent starting point in the process of
understanding these more complicated systems.  Indeed, we will also
show here that all qualitative conclusions made for the gKPR scheme
also hold in numerical studies of more complicated systems in which
rates are site dependent and where the connections of the nodes are
much more varied than a simple linear chain.

\subsection{The Relevant Features}
To determine if a kinetic model can be well approximated by a simpler
one, we must first decide which of its features must be retained. To
illustrate this question, consider the activation of a signaling
cascade by an extracellular ligand (as represented in Fig.\ \ref{fig:Model}).
The ligand binding initiates the process, bringing it from state $i=0$
to state $i=1$.  With the exception of this transition, the
extracellular environment does not affect the process.  Similarly, the
downstream signaling pathways are only affected when the signaling
construct attains its fully activated state at $i=L$.  Thus, as far as
the rest of the cell is concerned, only the times of process
initiation and completion are controllable, observable or otherwise
important. That is, the system can be characterized by the
distribution of the {\em first passage} or the {\em escape time}
between the release at $i=0$ at $t=0$ and the completion at $i=L$
\cite{Redner:2001}.  Analysis of this distribution and showing its
very simple limiting behavior is the main contribution of our work.

We note that, even though a lot is known about the first passage times
in different scenarios \cite{Redner:2001,Chou:2005,Shaevitz:2005} and
about temporal dynamics of KPR schemes
\cite{Ninio:1987,Liu:2001,Goldstein:2004}, to our knowledge, the
distribution of the first passage time for KPR type process has not
yet been analyzed rigorously and little is known regarding how this
first passage time depends upon biochemical parameters such as system
size and reaction rates.

\section{Results}

In the following subsections, we provide precise analyses of three
different cases of the gKPR scheme depicted in Fig.\ \ref{fig:Model},
each corresponding to a different continuous time / discrete space
Markov chain with exponential transition times (our results can be
generalized to the case of non-exponentially distributed transition
times using the methods of \cite{Bel:2006}). First is a normal random
walk process (that is $\gamma=0$) with an absorbing boundary at $i=L$
and a reflecting boundary at $i=0$. This model is denoted as the
transmission mode (TM) process \cite{Redner:2001}. The second model is
the directed KPR (dKPR) scheme where $(k>0, r=0, \gamma>0)$. The third
model is the full gKPR process, where all rates are non-zero.  For
each model, we provide exact solutions for the escape time
distributions in the Laplace domain and explicit expressions for the
mean and variances of the escape times (see also derivations in {\em
  Materials and Methods}).  By considering the squared coefficient of
variation, ${\rm CV}^2$, for these processes (see Figs.\ 3 and 5), we
explore how these distributions change as the system parameters are
adjusted and expose the fact that all three processes exhibit similar,
yet not identical, behavior. In particular, we find that all three
processes exhibit sharp transitions from near-deterministic
(${\rm CV}^2\ll 1$) to exponential (${\rm CV}^2=1$) completion times
as the critical parameters change, but that the actual location of
this transition differs between the TM and KPR processes. Furthermore,
we observe that all these processes have the same limiting behaviors
on either side of the transition, and that the transition from one
behavior to the other becomes sharper as the system size increases.
Finally, in Subsections 2.5 and 2.6, we also numerically explore the
same first passage time properties for more complicated cases where
the reaction rates are site dependent, and where more complicated
reaction events are possible.  For these processes, we again observe
the same simplifying behavior in the process dynamics and sharp
transitions that depend on the size of the system (see Figs.\ 8 and
9).

\subsection{Transmission Mode (TM)}
For the TM process, in which the forward and backward rates ($k$ and
$r$) are non-zero, one can derive explicit expressions for the mean
and the variance of the first passage time (see {\em Materials and
  Methods: Transmission Mode}).  Defining $\theta=r/k$, these can be
written:
\begin{eqnarray}
  \mu_{\rm TM}&=\frac{1}{k}\frac{L - (L+1) \theta+\theta^{L+1}}
  {(1-\theta)^2},\label{TTM} \\
  \sigma^2_{\rm TM}&= {\rm CV}^2_{\rm TM}\, \mu_{\rm TM}^2,\label{VTM}\\
  {\rm CV}^2_{\rm TM} & =\frac{L-4\theta-
    \left(L+1\right)\theta^2+4\left(L-L\theta +1\right)\theta^{L+1}+\theta^{2L+2}}
  {\left(L-L\theta+\theta\left[\theta^L-1\right]\right)^2},
  \label{CV_TM}
\end{eqnarray}
where ${\rm CV}_{\rm TM}$ is called the {\em coefficient of
  variation}. For a deterministic process, ${\rm CV}=0$, and for an
exponentially distributed one, ${\rm CV}=1$. This makes the
coefficient of variation a useful property characterizing a
distribution.

\begin{figure}[t]
\begin{center}
\includegraphics[width=5.5in]{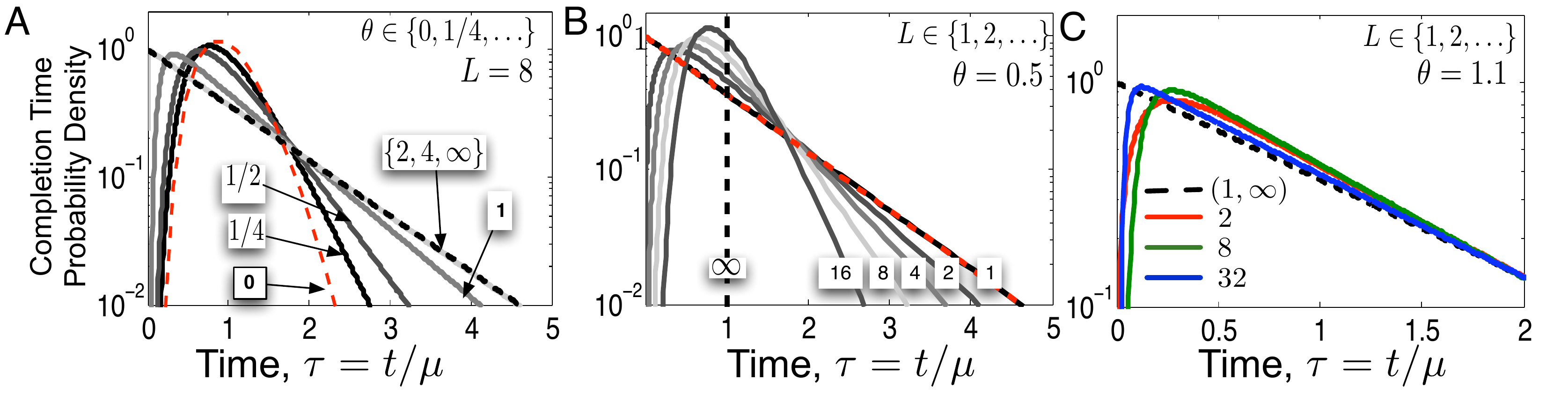}
\end{center}
\caption{{\bf Effect of changing $\theta=r/k$ and $L$ on the first passage
  time distribution for the TM process.} The time been rescaled for
  each curve as $\tau=t/\mu$.  (A) First passage time distribution
  for different values of the backward rate, $r$, and a
  fixed length $L=8$.  Here $r$ ranges from $k/4$ to $4k$, as denoted
  in the boxes.  The two dashed lines correspond to the limiting
  cases, $\theta=0,\infty$ ($\Gamma$-distribution and an exponential,
  respectively). (B,C) Effect of changing the length $L$ on the escape
  time distribution (B) for $\theta=0.5$ and (C) for $\theta=1.1$.
  For $\theta<1$, the limiting behavior as $L\rightarrow \infty$ is a
  delta function; for $\theta>1$, the limiting distribution is
  the exponential. \label{fig:PDF_v_r}}
\end{figure}

Fig.\ \ref{fig:PDF_v_r}A-C shows the effects that changes in the
parameters $\theta$ and $L$ have on the distribution of the escape
time.  In order to show the distribution for diverse parameters
simultaneously, time has been rescaled by the mean $\mu$ for each
curve, $\tau=t/\mu$.  This leads to the probability density
$f(\tau)=\mu f(t)$.  Fig.\ \ref{fig:PDF_v_r}A shows that, for a fixed
$L$, as $\theta$ increases, the distribution becomes broader and
approaches an exponential distribution, while as $\theta$ decreases
the distribution approaches a $\Gamma$-distribution,
$\Gamma(L,1/k)$. In order to quantify these behaviors we provide the
trends of the mean and the coefficient of variation for the
corresponding regimes.
\begin{eqnarray}\label{lim_mu_TM}
  \mu_{\rm TM}(L,\theta) &\approx 
  \left\{ \begin{array}{ll} 
      \theta^{L-1}/k &\textrm{ for $\theta \gg 2$,} \\
      L/k &\textrm{ for $\theta \ll \frac{L}{L-1}$,}
\end{array}\right.\\
\label{lim_CV_TM}
  {\rm CV}_{\rm TM}^2(L,\theta) &\approx 
  \left\{ \begin{array}{ll} 
      1-2(L-1)/\theta^L &\textrm{ for $\theta \gg \frac{L+2}{L-1}$,} \\
      1/L &\textrm{ for $\theta  \ll \frac{L}{2(L-1)}$.}
\end{array}\right.
\end{eqnarray}
It is worth mentioning that $\theta=1$ means an unbiased random walk,
while $\theta<1 (>1)$ means a walk biased towards the entry (exit)
point.

Figs.\ \ref{fig:PDF_v_r}B, C show that changes in $L$ have different
effects on the escape time distribution depending upon the value of
$\theta$.  When $\theta<1$, the limiting distribution as $L$ becomes
large is a $\delta$-function at $t = L/[k(1-\theta)]$, whereas for
$\theta>1$, the limiting distribution is an exponential with $\mu_{\rm
  TM}=\theta^{L+1}/[k(1-\theta)^2]$.

\begin{figure}[t]
\begin{center}
\includegraphics[width=10cm]{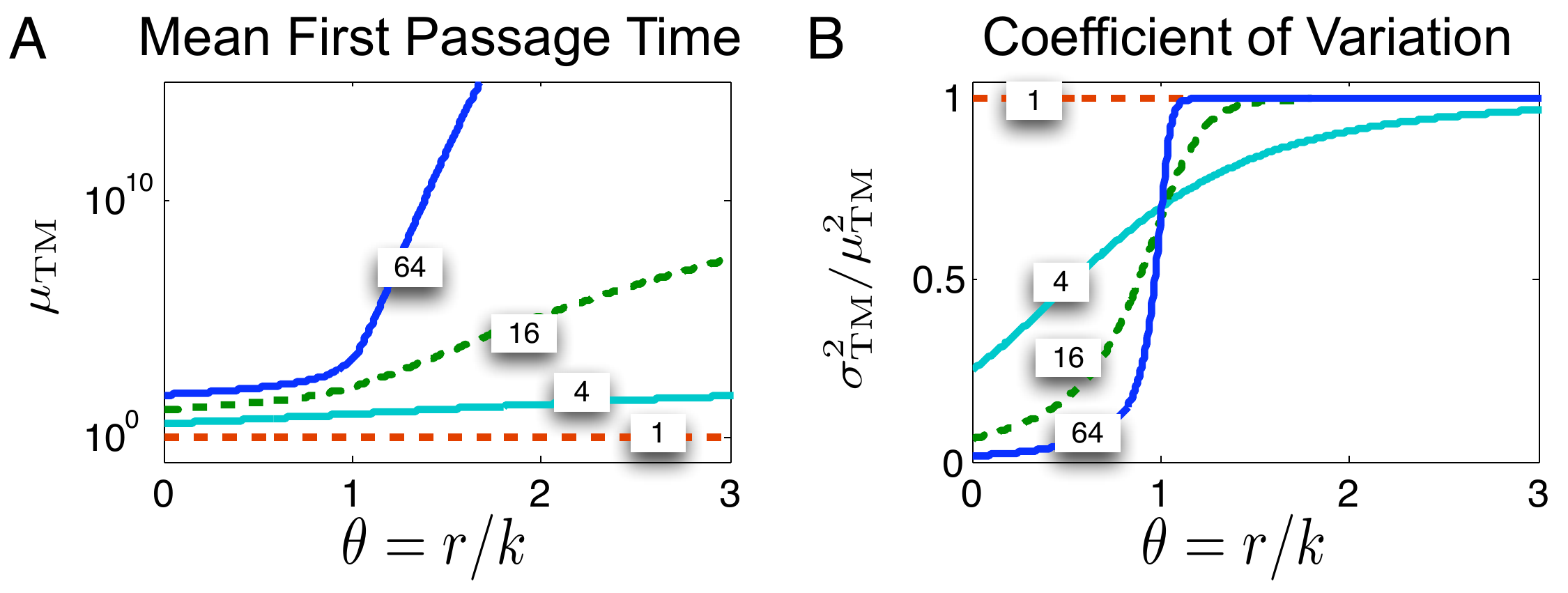}
\end{center}
\caption{{\bf Effect of changing the length and backward rate, $r$, on
    the mean (A) and squared coefficient of variation (B) of the TM
    process first passage times.}  The curves have been computed using
  Eqns.~(\ref{TTM}, \ref{CV_TM}) and are plotted for increasing values
  of $L=\{1,2,4,8,16,32\}$. \label{fig:MN_v_r}}
\end{figure}

Fig.\ \ref{fig:MN_v_r} illustrates the effect that changes in $L$ and
$\theta$ have on $\mu_{\rm TM}$ and ${\rm CV}^2_{\rm TM}$, as given by
Eqns.~(\ref{TTM},~\ref{CV_TM}).  It is of particular interest to
examine these as the chain becomes long. From Eq.~(\ref{CV_TM}), we
see that, as $L$ increases, ${\rm CV}_{\rm TM}^2$ converges point-wise
to the step function:
\begin{equation}
  \lim_{L\rightarrow \infty} {\rm CV}_{\rm TM}^2(L,\theta) =u(\theta-1)= 
  \left\{ \begin{array}{ll} 
      0 &\textrm{ for $\theta< 1$,} \\
      1 &\textrm{ for $\theta> 1$.}
\end{array}\right.
\end{equation}
Numerical analysis of Eq.~\ref{CV_TM} around $\theta=1$, shows that
the maximum slope of ${\rm CV}_{\rm TM}^2$ (to leading order in $L$)
occurs at a point that approaches $\theta=1$ at a rate:
\begin{equation}\label{t_pnt_TM}
1- \arg\max_{\theta} \frac{d{\rm CV}_{\rm TM}^2}{d\theta}=\frac{21}{2L^2}+\mathcal{O}(L^{-3}).
\end{equation}
The slope at $\theta=1-21/(2L^2)$ is:
\begin{equation}
  \max_{\theta} \frac{d{\rm CV}_{\rm TM}^2}{d\theta}=\frac{4}{45}L
  +\mathcal{O}(1). \label{cv_max_tm}
\end{equation}
Thus for a given large $L$, the range of $\theta$ over which the first
passage time changes from a narrow $\Gamma$-distribution to a broad
exponential distribution is centered just left of $\theta=1$, and it
becomes increasingly narrow as $L$ increases.

\subsection{Directed Kinetic Proofreading (dKPR)}
For the dKPR process, the system can return directly to the origin
with rate $\gamma>0$, but the backward rate, $r$, is zero. Then,
defining $\psi=\gamma/k$, the mean and the coefficient of variation of
the first passage times are (see {\em Materials and Methods: Directed
  Kinetic Proofreading}):
\begin{eqnarray}
  \mu_{\rm dKPR}&=\frac{1}{k\psi}\left[\left(1+\psi
    \right)^L-1\right], \label{TDKPR} \\
  {\rm CV}^2_{\rm dKPR}&=\frac{\left(1+\psi \right)^{2L}-2\psi
    L\left(1+\psi \right)^{L-1}-1}{(1+\psi
    )^{2L}-2(1+\psi)^L+1},\label{CV_DKPR}
\end{eqnarray}

\begin{figure}[t]
\begin{center}
\includegraphics[width=10cm]{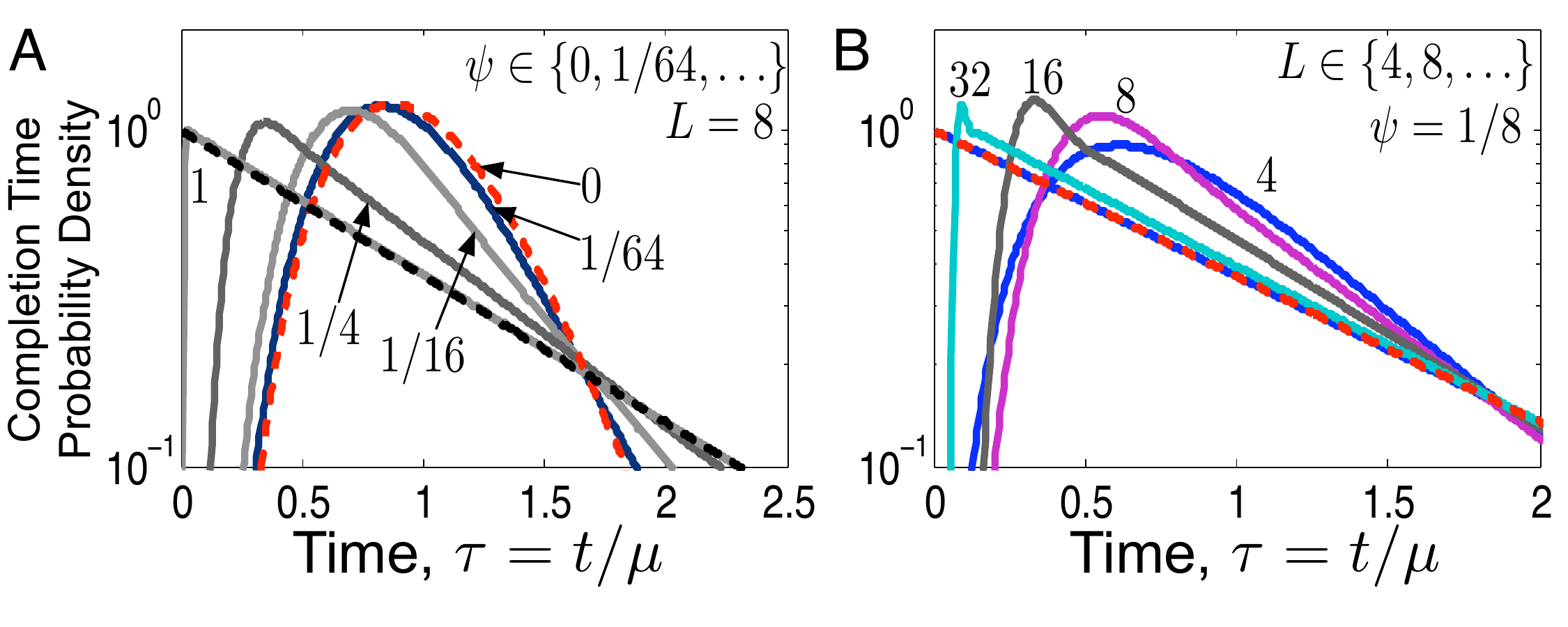}
\end{center}
\caption{{\bf Effect of changing $\psi=\gamma/k$ and $L$ on the first
    passage time distribution (normalized by its mean) for the dKPR
    process.}  (A) The first passage time distribution for different
  values of the return rate, $\gamma$ and a fixed length $L=8$.  The
  parameter $\psi$ ranges from $1/64$ to $1$ as denoted in the figure.
  The two dashed lines correspond to the limiting cases, where
  $\psi=0,\infty$.  The former results in a $\Gamma$-distribution, and
  the latter in an exponential distribution.  (B) Effect of changing
  the length $L$ on the first passage time distribution for
  $\psi=1/8$.  For any value of $\psi>0$, the limiting behavior as
  $L\rightarrow \infty$ is an exponential
  distribution. \label{fig:PDF_v_g}}
\end{figure}

Fig.\ \ref{fig:PDF_v_g}A-B shows the effects that changes in $\psi$
and $L$ have on the distribution of the waiting times for the dKPR
process.  As in the previous section, time has been rescaled by $\mu$
for each curve.  For a fixed $L$, as $\psi$ increases, the
distribution again approaches either an exponential distribution or
$\Gamma$-distribution for $\gamma\to\infty,0$, respectively.  Unlike
for the TM process, the limiting distribution as $L\to\infty$ is
exponential for any value of $\psi>0$.

\begin{figure}[t]
\begin{center}
\includegraphics[width=10cm]{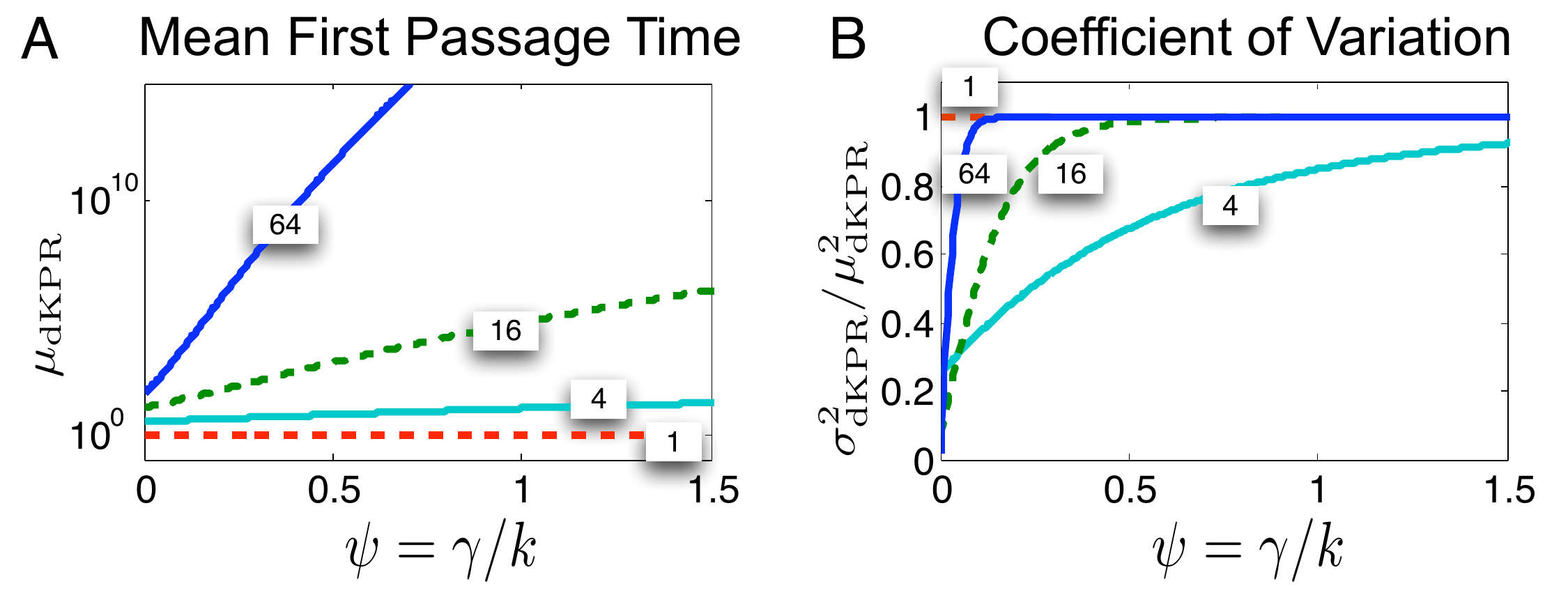}
\end{center}
\caption{{\bf Effect of changing the length and the proofreading rate,
    $\gamma$, on the mean (A) and the squared coefficient of variation
    (B) of the escape time for the dKPR system.}  The curves have been
  computed analytically using Eqns.~(\ref{TDKPR}, \ref{CV_DKPR}) and
  are plotted for increasing values of
  $L=\{1,2,4,8,16,32\}$.\label{fig:MN_v_g}}
\end{figure}

In Fig.\ \ref{fig:MN_v_g}, we illustrate the dependence of $\mu_{\rm
  dKPR}$ and ${\rm CV}_{\rm dKPR}^2$ on $L$ and $\psi$.  From Eqns.\
(\ref{TDKPR}, \ref{CV_DKPR}), their limiting behaviors are:
\begin{eqnarray}\label{lim_mu_dKPR}
  \mu_{\rm dKPR}(L,\psi) &\approx 
  \left\{ \begin{array}{ll} 
      \psi^{L-1}/k &\textrm{ for $\psi \gg L$,} \\
      L/k &\textrm{ for $\psi \ll L/2$,}
\end{array}\right.\\
\label{lim_CV_dKPR}
{\rm CV}_{\rm dKPR}^2(L,\psi) &\approx \left\{ \begin{array}{ll}
    1-{2(L-1)}/{\psi^L}&\textrm{for $\psi \gg 2L$,} \\
    {1}/{L}&\textrm{for $\psi \ll {3}/{L^2}$.}
\end{array}\right.
\end{eqnarray}
Furthermore, as $L$ grows, the coefficient of variation tends to
converge point-wise to a step function at $\psi=0$:
\begin{eqnarray}
  \lim_{L \rightarrow \infty} {\rm CV}^2_{\rm dKPR} =
  \left\{ \begin{array}{ll}
      0 &\textrm{ for $\psi=0$}, \\
      1 &\textrm{ for $\psi>0$}.
    \end{array}\right.
\end{eqnarray}

As in the TM process, this convergence can be studied by examining the
maximum slope of the coefficient of variation.  Since the second
derivative of ${\rm CV}^2_{\rm dKPR}$ is always negative for
$\psi\ge0$, this maximum slope occurs at $\psi=0$.  Taking the
derivative of Eqn. \ref{CV_DKPR} at the point $\psi=0$ yields an exact
expression for the maximal slope,
\begin{eqnarray}
  \max_{\psi} \frac{d{\rm CV}_{\rm dKPR}^2}{d\psi} &=\left.
    \frac{d{\rm CV}_{\rm dKPR}^2}{d\psi}\right|_{\psi=0}=\frac{L^2-1}{3L}.\label{cv_max_dkpr} 
\end{eqnarray}
These trends are readily apparent in Fig.\ \ref{fig:MN_v_g}B, where as
$L$ or $\psi$ increase, ${\rm CV}^2$ approaches unity. \newline

\subsection{Comparison between the TM and the dKPR models}
The TM and the dKPR processes exhibit very similar behaviors in their
first passage time distributions: for a fixed large $L$, increases in
$\theta$ or $\psi$ result in sharp transitions from deterministic to
exponential completion times. Moreover, the two processes have {\em
  quantitatively} the same limiting behaviors on either side of the
transition: the means and the CVs are asymptotically the same
functions of $\theta$ and $\psi$ [cf.\ Eqs.~(\ref{lim_mu_TM},
\ref{lim_CV_TM}, \ref{lim_mu_dKPR}, \ref{lim_CV_dKPR})].

However, the similarity between the limits of both processes is not
exact. For the TM, the deterministic-to-expo\-nent\-ial transition
(defined by the point of the maximum slope of ${\rm CV}^2$) is near
$\theta=1$, approaching it as $L$ grows [cf.\ Eq.~(\ref{t_pnt_TM})],
while the same transition for the dKPR is always at
$\psi=0$. Moreover, although for both models the width of the
transition region, as defined by the maximum slope of ${\rm CV}^2$, is
inversely proportional to the system size (for $L\gg1$), the width is
15/4 times larger for the TM process.  Finally, while the small/large
$\theta$ and $\psi$ limits are the same in both models, the terms {\em
  small} and {\em large} themselves have different meanings. In
particular, for the TM model the meanings are effectively independent
of the system size (Eqn.~\ref{lim_CV_TM}), while for the dKPR model
the meanings strongly depend on $L$ (Eqn.~\ref{lim_CV_dKPR}).

\subsection{General Kinetic Proofreading (gKPR)}
\begin{figure*}[t]
\begin{center}
\includegraphics[width=5.5in]{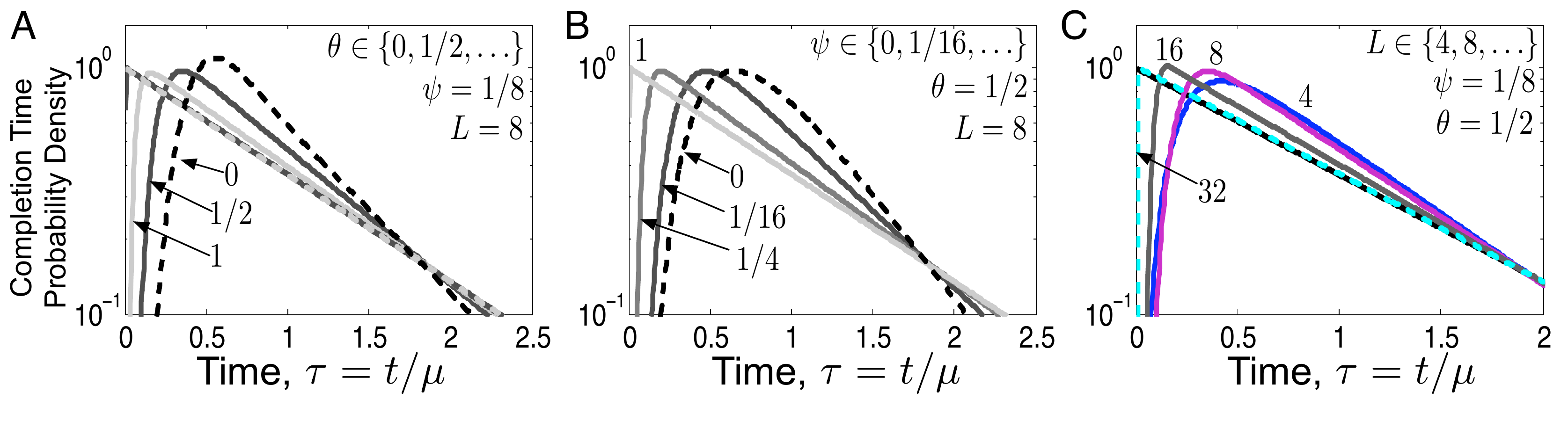}
\end{center}
\caption{{\bf The escape time probability density function for the
    gKPR scheme.}  (A) $\psi=\gamma/k=1/8$, $L=8$, and variable of
  $\theta=r/k$.  (B) $\theta=1/2$, $L=8$ and variable $\psi$. (C)
  $\theta=1/2$, $\psi=1/8$, and variable $L$. In all cases, the
  limiting behavior is an exponential as $L$, $\theta$, or $\psi$
  grow.}\label{fig:GKPR_PDF}
\end{figure*}

In the most general case, both $r>0$, and $\gamma>0$. Still, one can
derive explicit expressions for the mean and variance of the first
passage times (see Eqns.\ \ref{TGKPR}, \ref{VGKPR} in {\em Materials
  and Methods}).  Fig.\ \ref{fig:GKPR_PDF} illustrates the probability
distribution for the exit times of the gKPR process for different
$\theta$, $\psi$, and $L$.  Based upon the previous results, it is no
surprise that the escape time distributions converge to an exponential
distribution as $\psi$ or $\theta$ are large (cf.~Fig.\
\ref{fig:GKPR_PDF}A, B), or to a $\Gamma$-distribution when
$\psi=\theta=0$.  It is also not surprising that the gKPR first
passage time distribution converges to an exponential distribution
when $\gamma>0$ and $L$ is large (cf.~Fig.\ \ref{fig:GKPR_PDF}C).
What is surprising is how neatly the two constituent processes, TM and
dKPR, combine to define the trends of the gKPR process.

\begin{figure}[t]
\begin{center}
\includegraphics[width=10cm]{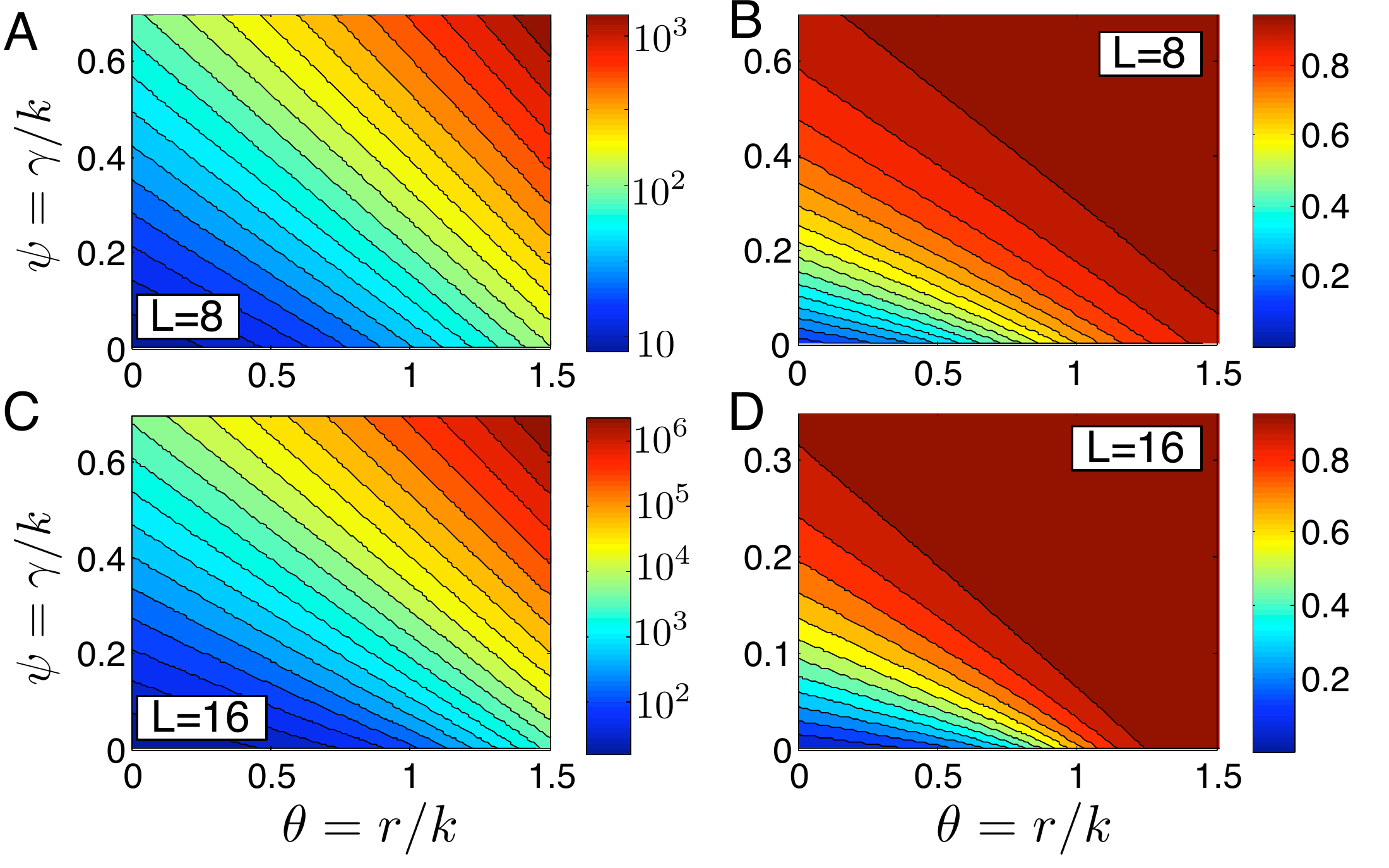}
\end{center}
\caption{{\bf Effects of parameter variation on the escape time
  distribution for the gKPR process.}  (A) Mean completion time
  versus $\theta$ and $\psi$ for $L=8$. (B) Coefficient of variation,
  ${\rm CV}_{\rm gKPR}^2$ versus $\theta$ and $\psi$ for $L=8$.  (C,
  D) the same for $L=16$.}\label{fig:GKPR_MN}
\end{figure}

Figs.\ \ref{fig:GKPR_MN}A-D show the mean and the coefficient of
variation of the first passage time distributions for this process
under various conditions.  In panel A, we plot $\mu_{\rm gKPR}$ as a
function of $\theta$ and $\psi$ for a fixed system size of $L=8$, and
panel B shows the corresponding ${\rm CV}_{\rm gKPR}^2$.  Panels C and
D show the same information, but for $L=16$.  We see that the general
trend for the increase in the mean passage time and the convergence of
the ${\rm CV}^2$ are determined in the same manner as those for the TM
and dKPR processes.  In particular, we find that that the contour
lines for both $\mu_{\rm gKPR}$ and ${\rm CV}^2_{\rm gKPR}$ are almost
linear.  However, this linearity is not exact--the actual contour
lines for $\mu_{\rm gKPR}(\psi,\theta)$ are slightly concave and the
contour lines for ${\rm CV}^2_{\rm gKPR}(\psi,\theta)$ are slightly
convex.  From Figs.\ \ref{fig:MN_v_r} and \ref{fig:MN_v_g} above, we
see that changes in $L$ have a large effect on the first passage time
of the TM and dKPR processes particularly around $\theta=1$ and
$\psi=0$, respectively.  In the gKPR process, these effects correspond
to changes in the endpoints, and therefore the slopes of the contour
lines in Fig.\ \ref{fig:GKPR_MN}A-D.

With explicit expressions for the mean and coefficient of variation,
one can again examine their limiting behaviors for growing $\psi$ and
$\theta$.  In particular, we find that these are equal to those of the
TM and the dKPR models when $\theta\to\infty$ or $\psi\to\infty$,
respectively.  Further, if $L$ is large and $\psi>0$, the mean first passage
is:
\begin{eqnarray}
  \lim_{L \rightarrow \infty} \mu_{\rm gKPR} &\approx
  \frac{(l_+\theta)^L}{2k\psi}\left(1+\frac{1-\theta+\psi}
    {\sqrt{(1+\theta+\psi)^2-4\theta}}\right),
\end{eqnarray}
where
\begin{equation}
l_+\theta =\frac{1+\theta+\psi+\sqrt{(1+\theta+\psi)^2-4\theta}}{2}
\geq 1.
\label{Lpm}
\end{equation}

Further, the coefficient of variation, ${\rm CV}_{\rm gKPR}^2$
approaches unity for all values except when $\psi=0$ {\em and}
$\theta<1$, and
\begin{equation}\label{lim_CV_gKPR}
  {\rm CV}_{\rm gKPR}^2(L,\theta) \approx 
  \left\{ \begin{array}{ll} 
      1-\frac{2(L-1)}{\left(\psi+\theta\right)^L} &\textrm{ for $\psi \gg 2L$ and $\theta \gg 4$,} \\
      1/L &\textrm{ for $\psi \ll \frac{L^2}{3}$ and $\theta \ll \frac{1}{2}$.}
\end{array}\right.
\end{equation}
This shows that, for large proofreading and backward rates, the two
effects have equal influences on the distribution of the completion
time. However, one should bear in mind that, again, the meaning of
small/large $\theta,\psi$ is different.

\subsection{Kinetic Proofreading with Site-Dependent Rates}
\begin{figure}[t]
\begin{center}
\includegraphics[width=5.5in]{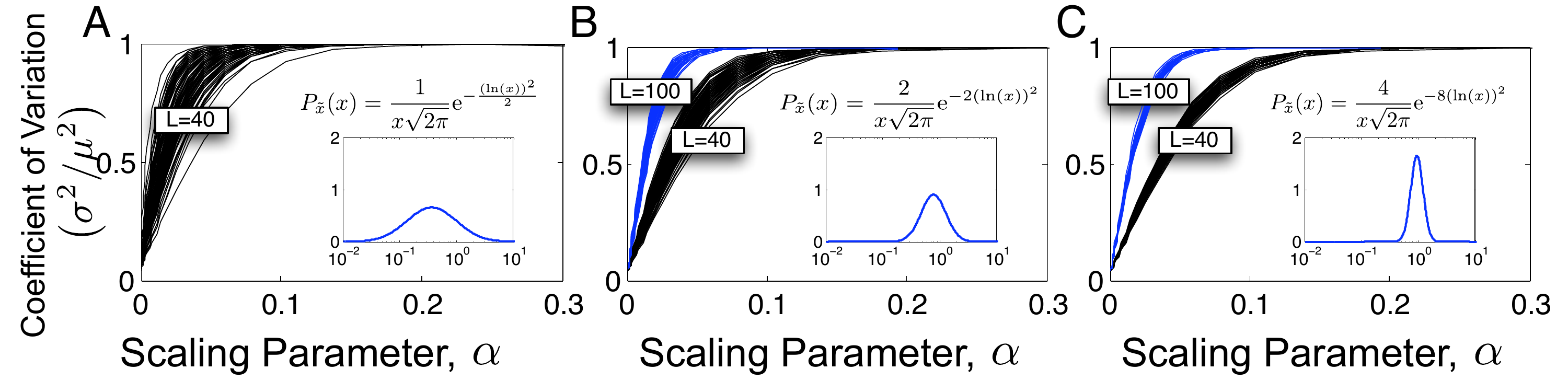}
\end{center}
\caption{{\bf Coefficient of variation for a gKPR process with random
    parameters versus backward to forward bias.}  The length of the
  process is either 40 (black lines) or 100 (blue lines) and all rates
  $k_i$, $\gamma_i/\alpha$ and $r_i/\alpha$ are taken independently
  from the lognormal distribution shown in the inset.  The three
  panels correspond to three increasingly narrow distributions for the
  parameters.\label{fig:SD_CV}}
\end{figure}

The previous subsections have shown that the TM, dKPR and gKPR
processes all exhibit a similar simplification of behavior when all
rates are the same at every intermediate state in the process.  In
reality, these rates may vary from one site to the next since each
transition may correspond to a different physical reaction.  In the
case of the dKPR, one can still derive expressions for the first
passage time distributions (see Materials and Methods), and in the
case of more complicated processes, one can explore these
distributions numerically.  To illustrate the effects of such
variation, we have numerically explored a gKPR process where every
rate is different, but chosen from some relatively broad lognormal
distribution.  Fig.\ \ref{fig:SD_CV} shows how such site dependent
rates affect the coefficient of variation for the gKPR process.  Here
all forward and backward rates, $\{r_i,k_i,\gamma_i\}$, have been
generated from the same distribution, and then the backward rates
$\{\gamma_i,r_i\}$ have been scaled uniformly by a parameter, $\alpha$
that has been used to adjust the bias from completely forward
$\alpha=0$ to backward $\alpha\gg 0$.  From Fig.\ \ref{fig:SD_CV} we
see once again that there is a sharp transition from when the
coefficient of variation is small at $\alpha=0$ to when the
coefficient of variation is near one when the bias is backward.  As in
the previous systems, this transition depends upon on the length of
the system---longer lengths correspond to sharper transitions.
Furthermore, as the lengths increase, variation in the parameters
appears becomes less important as can be seen by the comparing the
variation in the curves corresponding to $L=100$ (blue curves) to those
for a smaller length of $L=40$ (black curves).

\subsection{Multiple Leap Completion Processes}
In addition to the gKPR scheme illustrated by Fig.\ \ref{fig:Model},
we also explore a much more general set of multistep completion
processes where reactions can take the system not just one, but many
steps toward the completion state or toward the initial state.  In
terms of chemical processes, these multiple step jumps could
correspond to additions or removals of different multi-molecular
complexes rather than just individual molecules.  In this case there
are now many different interconnected pathways by which the process
can travel from state $i=0$ to $i=L$. In such systems, the master
equation, $d\mathbf{P}/dt = \mathbf{A}\cdot \mathbf{P}(t)$, has an
infinitesimal generator, $\mathbf{A}$ given by $\mathbf{A}=\alpha
\mathbf{B}+ \mathbf{F}$, where the ``backward'' matrix, $\mathbf{B}$ is
upper-triangular and represents reactions that allow the system to
return an arbitrary number of states backwards with certain
site-dependent rates, and the ``forward'' matrix $\mathbf{F}$ is a
lower-triangular banded matrix, which allows for different forward
jumps of lengths $m<L$, again with site-dependent rates.  Since $m$ is
constrained to be less than $L$, there is always a minimum of about
$L/m$ jumps necessary to complete the process. 

\begin{figure}[t]
\begin{center}
\includegraphics[width=5.5in]{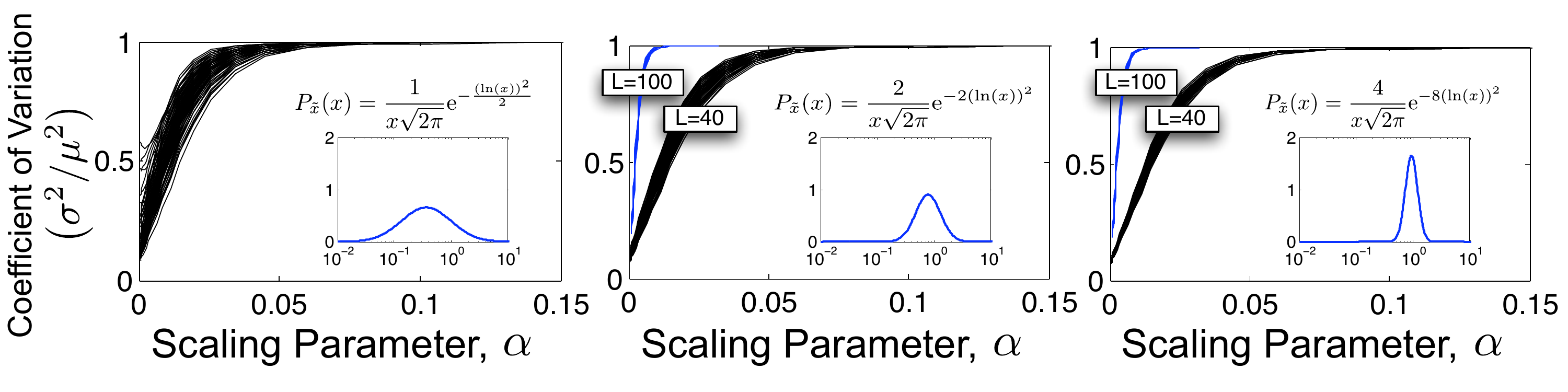}
\caption{{\bf Coefficient of variation for an arbitrary kinetic
    proofreading like process with random parameters versus backward
    to forward bias.}  The master equation for this process is
  $\dot{\mathbf{P}}(t)=\left(\alpha\mathbf{B} +
    \mathbf{F}\right)\mathbf{P}(t)$, where $\mathbf{F}$ is banded such
  that the system can move 1, 2, or 3 steps forward in a single jump,
  and $\mathbf{B}$ is upper triangular such that the process can move
  any number of steps backwards.  The length of the process is either
  40 (black lines) or 100 (blue lines), and each non-zero element of
  $\mathbf{F}$ and $\mathbf{B}$ is randomly chosen from the lognormal
  distribution plotted in the inset. The three panels correspond to
  three increasingly narrow distributions for the
  parameters.\label{fig:ARB_LU}}
\end{center}
\end{figure}

In the expression of the infinitesimal generator, $\alpha$ controls
the bias, and we show once again that there is a sharp threshold
between an almost deterministic and an exponential behavior as
$\alpha$ grows.  For this arbitrary process, we have randomly
generated hundreds of realizations each with different site-dependent
rates taken from a broad lognormal distribution, and we find that for
every such parameter set, there is a sharp transition from a narrow
``deterministic'' to a broad exponential waiting time distribution as
can be seen in Fig.\ \ref{fig:ARB_LU}.  Furthermore, despite drastic
differences in the randomly chosen parameters, we find that the
dynamical behaviors of the systems are so close that it is difficult
to distinguish one parameter set from the next based solely on the
waiting time.  Finally, we find the same dependence of this transition
on the size of the system as has been observed for dKPR and gKPR
processes (compare the process with 40 steps (black lines) to the
process with 100 steps (blue lines) in Fig.\ \ref{fig:ARB_LU}).

\section{Discussion}
The results for the coefficient of variation of the escape time
distribution, as well as the shapes of the distributions themselves,
clearly show that the kinetic proofreading process and other multistep
completion processes have two simple limiting behaviors as the system
size increases. First, when the overall bias is forward, the
completion time becomes narrowly distributed. Second, when the overall
bias is backward, the escape time distribution approaches an
exponential. Both of these behaviors are substantially simpler than
one could have expected from the the original complex kinetic diagram,
implying that the observable behavior of this complex system can be
approximated accurately by a single-parameter equivalent,
corresponding either to a deterministic reaction or a simple two-state
Markov chain.  Interestingly, the approach to the deterministic regime
as the system size grows is well understood (see, for example,
\cite{doan-06} on the discussion of this effect in the context of
reproducibility of responses of rod cells to single photon capture
events). However, the exponential regime has not been explored
extensively before, even though it is the more robust of the two,
emerging for any $\psi>0$.

Both limiting behaviors of these systems are explainable by simple
intuitive arguments. First, a system with a forward bias completes the
entire process in a certain characteristic time, and the relative
standard deviation of this time scales as $1/\sqrt{\textrm {number of
    steps}}$, as is always the case for the addition of independent
identically distributed random variables. In the opposite case, the
backward bias ensures that the process repeatedly returns to the
initial state, from which many {\em independent} escape attempts are
made. Due to the independence, the number of such attempts before a
success has a geometric distribution (the discrete analog of an
exponential distribution), and its form effectively defines the first
passage time distribution. In other words, the system tries to climb
out of a free energy well (with the ground state near the entry
point), and escape times in such cases are usually exponentially
distributed.

Although the KPR models most rigorously analyzed here are relatively
simple linear chain processes with site-independent transition rates,
our numerical studies strongly suggest that the conclusions we make
generalize to more complicated systems.  We have shown numerically
that our conclusions do not change when the kinetic rates $k,r,\gamma$
are site-specific and/or when the reactions allow for certain states
to be skipped and for there to be many different interconnected
pathways by which the process may be completed.
Similarly, if biochemical processes involve multiple independent
pathways, each with exponential/deterministic waiting times, then the
first of these pathways to complete will also be
exponential/deterministic.  Furthermore, first passage times for
higher dimensional random walks also frequently exhibit simplified
dynamics, as has been shown via reductions to a stochastic model of
the genetic toggle switch \cite{Munsky:2008IET}.  Finally, the ``free
energy well'' argument says that the overall bias of a system's motion
will control the choice between the exponential (Markovian) and the
deterministic behaviors even for more complex systems.  In particular,
it is clear that any KPR-like system, where a strong backward bias is
required to undo potential mistakes, is likely to fall in the
exponential escape time distribution regime.

Given that so much structural complexity is used to achieve a very
simple dynamics in these processes, it is natural to ask why the
complexity is used at all. One hypothesis is that such agglomeration
of multiple independent kinetic parameters into a few coarse-grained
variables means that multiple chemotypes can result in the same
phenotype. Thus, the system possesses {\em many} situationally {\em
  sensitive} knobs with which it can compensate for environmental
changes and maintain {\em a few} simple behaviors. Such adaptive
flexibility has been observed in a variety of contexts
\cite{braun-07,ziv-etal-07,sloppy}.  An alternative hypothesis may be
that these extra elements are vestigial network components to which
the cell is {\em insensitive} in its current evolutionary or
developmental situation.  The current work provides a starting point
to evaluate these possibilities via parametric sensitivity analysis.
 
Finally, the fact that the KPR process, as well as many others, has
such simple limiting behaviors has important consequences for the
modeling of biochemical systems.  The bad news is that it is
unreasonable to hope to characterize individual molecular reactions
with observations of the input-to-output responses---many different
internal organizations will result in equivalent observable behaviors.
The good news is that, when attempting to understand such processes in
a wider cellular context, it is often unnecessary to explicitly treat
every individual step--a coarse-grained model with only a handful of
aggregate parameters may be sufficient. This result clearly explains
why simple phenomenological Markovian reaction rate models of
complicated processes, such as transcription, translation, enzyme
activation and others, have had such a great success in explaining
biological data.

\section*{Materials and Methods}
\subsection*{Preliminaries}
Let the vector $\mathbf{p} = [p_0(t), p_1(t), \ldots, p_{L}(t)]^T$
denote the probabilities of each state in the kinetic diagram in
Fig.~\ref{fig:Model}. This distribution evolves according to the
Master Equation (ME), which can be written: $\dot{\mathbf{p}}(t) =
\mathbf{Ap}(t)$, where the infinitesimal generator matrix $\mathbf{A}$
is:
\begin{equation}
A_{ij} = \left\{ \begin{array}{ll} 
-k & \textrm{for $i=j=0$}, \\ 
-k-\gamma-r &\textrm{for $0<i=j\leq L-1$}, \\ 
\gamma+r &\textrm{for $(i,j) = (0,1)$}, \\ 
\gamma &\textrm{for $i=0$ and $2\leq j \leq L-1$}, \\ 
r &\textrm{for $i=j-1$ and $2\leq j \leq L-1$}, \\ 
k &\textrm{for $i=j+1$ and $2\leq j \leq L-1$}, \\
0 &\textrm{everywhere else}.
\end{array} \right.
\end{equation}

By applying the Laplace transform,
\begin{equation}
  P_{i}(s)=\displaystyle{\int\limits_{0}^{\infty}}p_{i}(t) e^{-st}dt,
\end{equation}
one can convert the ME to a set of linear algebraic equations:
\begin{eqnarray}
  \left(s -\mathbf{A}\right)\mathbf{P}(s) =
  \mathbf{p}(t=0)\equiv\mathbf{e}_0.  \label{MELaplace}
\end{eqnarray}
Note that this equation includes the specification of the initial
condition, $p_{i}\left(t=0\right)=\delta_{i,0}$, where $\delta$ is the
Kronecker delta. 

We now construct a general solution for this equation in the form
\begin{equation}
  P_{i}(s) =C_1\lambda _{1}^{i}+C_2\lambda _{2}^{i}. \label{GenSol}
\end{equation}
Inserting this into the expression for $0<i<L-1$, one finds that the
space-independent parameters $\lambda_{1,2}$ satisfy
\begin{equation}
\frac{k}{s+k+\gamma +r}+\frac{r}{s+k+\gamma +r}\lambda ^{2}-\lambda=0. \label{quad}
\end{equation}
Similarly, the coefficients $C_1$ and $C_2$ must obey the
equations for $P_0(s)$ and $P_{L-1}(s)$ in (\ref{MELaplace}), which
can be written as
\begin{eqnarray}
  \left(s+k\right)\left(C_1+C_2\right)&=
  1+r\left(C_1\lambda_1+C_2\lambda_2\right)+\nonumber\\
  &\gamma\left(C_1\left[\frac{1-\lambda_{1}^{L}}{1-\lambda_1}-1\right]
    +C_2\left[\frac{1-\lambda_{2}^{L}}{1-\lambda_2}-1\right]\right) \\
  C_1\lambda_{1}^{L-1}+C_2\lambda_{2}^{L-1}&=
  \frac{k}{s+k+r+\gamma}\left(C_1\lambda_{1}^{L-2}+
   C_2\lambda_{2}^{L-2}\right),\label{ABeqs}
\end{eqnarray}
where we have applied the geometric series identity,
$\sum_{i=1}^{L-1}\lambda^i = \frac{1-\lambda^{L}}{1-\lambda}-1$. 

Since $P_L(t)$ is the cumulative probability that the system has
reached the absorbing state, the first passage time probability
density, $f(t)=dp_L(t)/dt$, can be written in the Laplace domain as:
\begin{equation}
F(s)=kP_{L-1}(s). \label{LFPTPDF}
\end{equation}
Once this quantity is known, all uncentered moments of the escape
time are easily derived as 
\begin{equation}
  T^{(m)}=\displaystyle{\int\limits_{0}^{\infty}t^mf(t) dt}=\left. \left(-1\right)^m\frac{dF(s)}{ds}\right|_{s=0}. \label{moments}
\end{equation}
With this in mind, we now consider the three special cases in the
following subsections.

\subsection*{Transmission Mode}
The first case to be considered is transmission mode: the continuous
time, discrete space random walk, where the process can only move
forward or backward to its nearest neighbor.  Applying the boundary
conditions as expressed in Eq.~(\ref{ABeqs}) yields the expressions
for $C_1$ and $C_2$:
\begin{eqnarray}
  C_1&=\frac{1}{\left(s+k-r\lambda_2\right)
    \left[\frac{\lambda_2-1}{\lambda_1-1}- \left(
        \frac{\lambda_1}{\lambda_2} \right)^L \right]}, \;\mbox{and}\;
  C_2=-C_1\frac{\lambda _{1}^{L}}{\lambda _{2}^{L}}, \label{ABTM}
\end{eqnarray}
where $\lambda_1$ and $\lambda_2$ are obtained from Eq.~(\ref{quad}):
\begin{equation}
  \lambda_{1,2}=\frac{s+k+r \pm \sqrt{\left(s+k+r\right)^2-4kr} }{2r}. \label{lambdaTM}
\end{equation}

Following simple algebra, the Laplace transform of the first passage
time probability density function (PDF) then becomes
\begin{equation}
  F(s)=C_1 k\lambda_1^{L-1}\left(1-\frac{\lambda_1}{\lambda_2}\right), \label{FPTPDFTM}
\end{equation}
from which all moments of the first passage time can be extracted. In
particular the mean escape time and the variance are given by
Eqs.~(\ref{TTM}, \ref{VTM}) in the main text.

\subsection*{Directed Kinetic Proofreading}
The second case we consider is that of directed kinetic proofreading,
in which the backward transition rate is neglected, $r=0$, but the
return rate is non-zero, $\gamma>0$. In this case the solution is much
simpler and can be written as
\begin{equation}
\tilde{p}_{i}(s) =C_1\lambda^{i}, 
\end{equation}
where $\lambda$ is the single root of Eq.~(\ref{quad}) given by
\begin{equation}
\lambda=\frac{k}{s+k+\gamma}, \label{quadDKPR}
\end{equation}
and the coefficient $C_1$ is reduced to
\begin{eqnarray}
  C_1=\frac{1}{s+k-\gamma\left(\frac{1-\lambda^L}{1-\lambda}-1\right)}. \label{ABeqsDKPR}
\end{eqnarray}
In this case, the Laplace transform of the first passage time is given
by
\begin{equation}
  f(s)=kp_{L-1}(s)=\frac{k}{s+k-
    \gamma\left(\frac{1-\lambda^L}{1-\lambda}-1\right)}\lambda^{L-1}, \label{FPTPDFDKPR}
\end{equation}
which gives the expressions for the mean escape time and its
coefficient of variation as in Eqs.~(\ref{TDKPR}, \ref{CV_DKPR}) in
the main text.  In the case of site-dependent rates, one can still derive an expression for the Laplace transform of the completion time, which can be written as:
\begin{equation}
f\left(s\right)=\frac{k_{L-1}\displaystyle{\prod_{j=1}^{L-1}}\frac{k_{j-1}}{s+k_{j}+\gamma_{j}}}{s+k_0-\displaystyle{\sum_{i=1}^{L-1}}\gamma_{i}\displaystyle{\prod_{j=1}^{i}}\frac{k_{j-1}}{s+k_j+\gamma_j}}.
\end{equation}

\subsection*{General Kinetic Proofreading}
In this case, all the rates $k$, $\gamma$, and $r$ are non-zero, and
Eq.~(\ref{quad}) has two solutions
\begin{equation}
  \lambda_{1,2}=\frac{s+k+r+\gamma\pm\sqrt{\left(s+k+r+\gamma\right)-4kr}}{2r}. \label{quadGKPR}
\end{equation}
By applying the boundary conditions in Eq.~(\ref{ABeqs}), we obtain
the expressions for $C_1$ and $C_2$:
\begin{eqnarray}
  C_1&=\frac{1}{r\left(\lambda_2-1\right)-
    \gamma\frac{1-\lambda_1^L}{1-\lambda_1}
    +\left(\frac{\lambda_1}{\lambda_2}\right)^L
    \left(r\left(1-\lambda_1\right)+
      \gamma\frac{1-\lambda_2^L}{1-\lambda_2}\right)} \\
  C_2&=-C_1\left(\frac{\lambda_1}{\lambda_2}\right)^L, \label{ABeqsGKPR}
\end{eqnarray}
with which one can define the Laplace transform of the first passage
time PDF:
\begin{equation}
  F(s)=C_1 k\lambda_1^{L-1}\left(1-\frac{\lambda_1}{\lambda_2}\right). \label{FPTPDFGKPR}
\end{equation}
Once again, it is possible to derive the the mean and variance of the
escape time in this scheme
\begin{equation}
  \mu_{\rm gKPR} =\frac{1}{2k\psi}\left[\frac{1-\theta+\psi}
    {\sqrt{\left(1+\theta+\psi\right)^2-4\theta}}
    \left(l_{+}^L-l_{-}^L\right)\theta^L
  +\left(l_{+}^L+l_{-}^L\right) \theta^L-2\right], \label{TGKPR}
\end{equation}
where $l_{\pm}$ are defined as in Eq.~(\ref{Lpm}).  The first passage
time variance in this case is given by
\begin{eqnarray}
  k^2\psi^2\sigma^2_{\rm gKPR}&= \frac{1}{2}\theta^{2L}
  \left(l_{-}^{2L}+l_{+}^{2L}\right)-1  \nonumber\\
  &+ \frac{\theta^{2L-1}\left(\theta - 1 - \psi \right)
    \left(l_{-}^{2L}-l_{+}^{2L}\right)+2L\psi\theta^{L-1}
    \left(l_{-}^L-l_{+}^L\right)}{2\left(l_{+}-l_{-}\right)}  \nonumber\\
  &+\psi\frac{2\theta- L\left(l_{-}^L+l_{+}^L\right) \left(-\theta +1 +
      \psi \right) 
    \theta^{L-2}- \theta^{2L-1}\left(l_{-}^{2L}+l_{+}^{2L}\right)}
  {\left(l_{+}-l_{-}\right)^2}  \nonumber \\
  &-\frac{2\psi\left(1 - \theta^{L-1} \right)}{ \left(l_+ -
      l_-\right)^2}
  +\frac{2\theta^{L-2} \psi
    \left(\theta - 1 + \psi \right) \left(l_{-}^L-l_{+}^L\right)}{\left(l_{+}-l_{-}\right)^3}. \label{VGKPR}
\end{eqnarray}

\section*{Acknowledgments}
We thank N.\ Sinitsyn and N. Hengartner for discussions during early stages of
  this work. We also thank B.\ Goldstein, R.\ Gutenkunst, and M.\
  Monine. This work was partially funded by LANL LDRD program.

\section*{References}
\bibliographystyle{IEEEtran}
\bibliography{KPR}

%


\end{document}